\documentclass{aa}
\usepackage{psfig}

\def \deg{^\circ}

\def \hcm {\hbox {\ifmmode $ cm$^{-2}\else cm$^{-2}$\fi}}
\def \arcmin {\hbox{$^\prime$}}
\def \arcsec {\hbox{$^{\prime\prime}$}}
\def \cmdue{\hbox{${\rm cm^{-2}}$}}
\def\approxgt{\mathrel{\hbox{\rlap{\lower.55ex \hbox {$\sim$}}
        \kern-.3em \raise.4ex \hbox{$>$}}}}
\def\approxlt{\mathrel{\hbox{\rlap{\lower.55ex \hbox {$\sim$}}
        \kern-.3em \raise.4ex \hbox{$<$}}}}

\begin{document}


\title{{1RXS J214303.7+065419/RBS 1774: \\
A New Isolated Neutron Star Candidate}
\thanks{Partially based on observations carried out at ESO, La Silla,
Chile (67.H--0116(A))}}

\author{L. Zampieri \inst{1,2}
        \and S. Campana \inst{3}
        \and R. Turolla \inst{2}
        \and M. Chieregato \inst{2},\\
             R. Falomo \inst{1}
        \and D. Fugazza \inst{3}
        \and A. Moretti \inst{3}
        \and A. Treves \inst{4}
}

\offprints{L. Zampieri (zampieri@pd.astro.it)}

\institute{
Osservatorio Astronomico di Padova, Vicolo dell'Osservatorio 5, I-35122
Padova, Italy
\and
Universit\`a di Padova, Dipartimento di Fisica, Via Marzolo 8, I-35131
Padova, Italy
\and
Osservatorio Astronomico di Brera, Via Bianchi 46, I-23807 Merate, Italy
\and
Universit\`a dell'Insubria, Dipartimento di Scienze, Via Valleggio 11,
I-22100 Como, Italy
}
\date{Received  ; Accepted  }

\markboth{A new isolated neutron star candidate}{A new isolated neutron
star candidate}

\abstract{We report on the identification of a new possible Isolated Neutron
Star candidate in archival ROSAT observations. The source
1RXS J214303.7+065419, listed in the ROSAT Bright Survey as RBS 1774,
is very soft, exhibits
a thermal spectrum well fitted by a blackbody at ${\rm T}\sim 90$ eV
and has a low column density, ${\rm N_H}\sim 5\times 10^{20}$ ${\rm
cm}^{-2}$.  Catalogue searches revealed no known sources in other
energy bands close to the X-ray position of RBS 1774. Follow-up
optical observations with NTT showed no peculiar object within the
X-ray error circle. The absence of any plausible optical counterpart
down to ${\rm m_R}\sim 23$ results in an X-ray to optical flux ratio
in excess of 1000.
}
\maketitle


\section{Introduction}

Over the last few years ROSAT observations
led to the discovery of six very soft X-ray sources with quite
peculiar characteristics. Among these are (i) blackbody-like
spectrum with ${\rm T}\sim 100$~eV; (ii) exceedingly
large X-ray to optical flux ratio, ${\rm f_X/f_V} > 10^3$; (iii) low X-ray
luminosity, ${\rm L_X}\approx 10^{30}-10^{31} {\rm erg\,s}^{-1}$; (iv)
low column density, ${\rm N_H}\sim 10^{20} \ {\rm cm}^{-2}$; (v) no evidence
for a binary companion; (vi) absence of large flux variations on
time-scales from months to years (see e.g. Treves et al. \cite{t2000};
Motch \cite{m2000} for reviews).

All these points, in particular the extreme values of ${\rm f_X/f_V}$
together with the small distances implied by the low column
density, qualify these sources as potential, close-by Isolated
Neutron Stars (INSs). Three sources have been found to pulsate
with periods in the range 5--23 s, strengthening further the
association with neutron stars (1RXS J130848.6+212708/RBS 1223, Haberl,
private communication; RX J0720.4-3125, Haberl et
al. \cite{hetal97}; RX J04020.0-5022, Haberl et al.
\cite{hetal99}). Intensive follow-up campaigns led
to the identification of an optical counterpart for
RX J1856.5-3754 (${\rm m_V}\sim 25.6$; Walter \& Matthews
\cite{wm97}; see also van Kerkwijk \&  Kulkarni \cite{kvk2001})
and, with less certainty, for RX J0720.4-3125 (${\rm m_B}
\approx 26$; Motch \& Haberl \cite{mh98}; Kulkarni \& van
Kerkwijk \cite{kk98}). Very recently, Walter (\cite{w2001}) obtained
a parallax distance of $\sim 60$ pc for RX J1856.5-3754 with the
Hubble Space Telescope, providing direct evidence that these sources
should indeed be located within a few hundred parsecs.

Although it is now widely agreed that the six ROSAT sources are
isolated neutron stars, their puzzling properties make the origin of their
emission still uncertain. Up to now, two
possibilities have been discussed: accretion of the interstellar medium
onto old INSs or thermal emission from much younger cooling neutron
stars. Due to the extreme similarity of the expected X-ray spectra, however,
present data still do not allow an unambiguous determination of the nature
of these sources (Treves et al. \cite{t2000}). Statistical methods
have been used to probe and compare
the distributions of cooling and accreting INSs (Ne\"uhauser \& Tr\"umper
\cite{nt99}; Popov et al. \cite{p2000}), but, being severely
hindered by the limited sample of detected sources, they fail to give
a definite answer. The similarity in the periods
has suggested a possible evolutionary
link between Anomalous X-ray Pulsars, Soft Gamma-ray Repeaters and ROSAT INSs
whereby the former may be the progenitors of the latter.
If this is the case, RX J0720.4-3125 might be a
magnetar kept hot by
the dissipation of a superstrong magnetic field (B$\approx 10^{14}-
10^{15}$~G; Heyl \& Hernquist \cite{hh98}; Heyl \& Kulkarni \cite{hk98}).

The discovery of new INS candidates is of paramount importance to
shed light on the properties of these sources and to constrain their
distribution in the Galaxy. In this paper we report the identification
of a new possible INS candidate in archival ROSAT observations.
X-ray and optical follow-up observations are described in Sect.~\ref{obser}.
Discussion on the nature of 1RXS J214303.7+065419/RBS 1774 and conclusions
are reported in Sect.~\ref{disc}.

\begin{figure}[!htb]
\centerline{\psfig{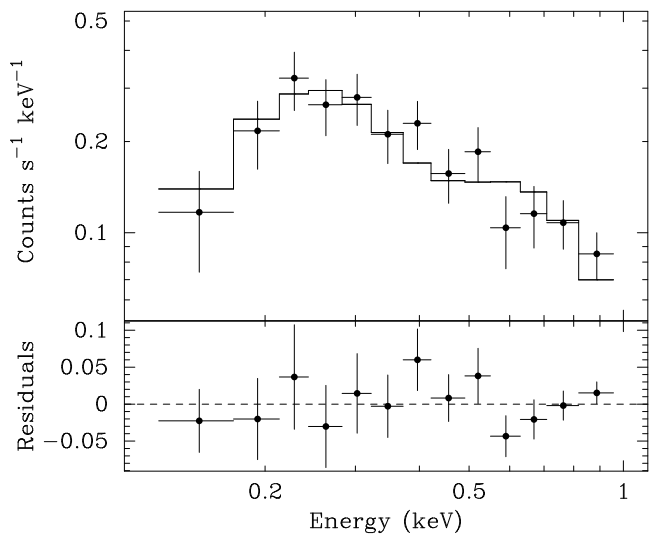}}
\caption[]{The ROSAT PSPC spectrum of RBS 1774 with the best-fitting
blackbody with ${\rm T}= 92$ eV and a low column
density, ${\rm N_H}=4.6\times 10^{20}$ ${\rm cm}^{-2}$.
\label{fig:spec}}
\end{figure}


\begin{table*}[htb]
\caption{X-ray spectral models.\label{fit_param}}
\begin{tabular}{lcccc}
Model          &Parameters$^1$     & ${\rm N_H}$              & $\chi^2_{\rm red}\, {\rm (dof)}$&  Flux$^2$ (0.1--2.4 keV)\\
               &              &($10^{20}\cmdue$)   &                  & (${\rm erg \, s^{-1} \, cm^{-2}}$)\\
\hline
Blackbody      &${\rm k\,T}=92^{+19}_{-15}\, {\rm eV}$    &$4.6^{+0.2}_{-0.2}$ & 0.83 (10)         &$8.7\times 10^{-12}$\\
Power law      &$\Gamma=5.5^{+1.2}_{-1.2}$ &$11.3^{+0.3}_{-0.3}$& 0.54 (10)         &$7.4\times 10^{-10}$\\
Raymond-Smith  &${\rm k\,T}=168^{+78}_{-41}\, {\rm eV}$, ${\rm Z<10^{-2}\,Z}_\odot$ &$7.3^{+0.2}_{-0.2}$ & 0.68 (9)&$3.0\times 10^{-11}$\\
Zampieri et al.&$\log({\rm L/L_{Edd}})=-7.2^{+0.8}_{-0.7}$&$6.9^{+0.3}_{-0.1}$ & 0.67 (10)         &$2.2\times 10^{-11}$\\
\end{tabular}

\smallskip
\noindent $^1$ Errors are at $90\%$ confidence level.
\smallskip

\noindent $^2$ Fluxes are unabsorbed.
\end{table*}

\section{Observations}
\label{obser}
\subsection{X-ray observations}
\label{xray-obs}

We re-examined archival ROSAT PSPC data looking for relatively
bright sources with ${\rm f_X/f_V} > 100$ and no likely
counterpart on Digital Sky Survey fields. The search on the ROSAT
Bright Survey (RBS; Schwope et al. \cite{s00})
yielded three sources with count rates $\approxgt 0.2$
counts$\,{\rm s}^{-1}$ matching the adopted criteria. Two of them
are the already known INS candidates RX J1605.3+3249/RBS 1556
(Motch et al. \cite{mo99}) and 1RXS J130848.6+212708/RBS 1223 (Schwope et al.
\cite{s99}). The remaining one, 1RXS J214303.7+065419/RBS 1774 (or
1WGA J2143.0+0655),
was serendipitously detected by ROSAT in a PSPC pointing of the BL Lac
object MSS 2143.4+0704 (Stocke et al. \cite{sto91}). 
The 3.5 ks observation was performed in May 1991
(ROSAT pointing \# rp700045n00). RBS 1774 is located about 48\arcmin \
off-axis, to the south-west of the position of MSS 2143.4+0704, and is
present both in the WGA (White et al. \cite{wga94}) and
the ROSATSRC (Zimmermann \cite{zim94}) catalogues. Examination of the
NED and SIMBAD astronomical databases did not show any catalogued
source within 5$\arcmin$ of the nominal RBS source position.

The analysis of the PSPC field provides a poor determination of the
source position because of the large off-axis angle. The nominal
positions in the WGA and ROSATSRC catalogues differ by $\sim
30\arcsec$. In the short (373 s) exposure taken during the ROSAT All
Sky Survey (RASS; Voges et al. \cite{vetal96}) the source is at a much
smaller average off-axis angle so, albeit few photons were collected,
it allows for a better position estimate. The RASS coordinates are
again different from those of the PSPC field. For this reason we
decided to run a simplified version of the Brera Multiscale Wavelet
(BMW) algorithm (Lazzati et al. \cite{laz99}; Campana et
al. \cite{cam99}) on the RASS field. The estimated position derived
from our re-analysis of RASS data is ${\rm RA} = 21^{\rm h}\, 43^{\rm
m}\, 02.0^{\rm s}$, ${\rm DEC} = +06\deg\, 54\arcmin\, 26\arcsec$
(J2000), with a $90\%$ uncertainty radius of $14\arcsec$.

Photons were extracted from a $6\arcmin\times 3.5\arcmin$
elliptical region  centered on the source position in the pointed
PSPC observation. A total of 763 photons were collected,
providing $\sim 500$ net counts after background subtraction. The
count rate is $0.18 \ {\rm counts\,s^{-1}}$. The spectral
analysis was carried out with XSPEC (v. 11.0.1).  Energy channels
were grouped into bins containing 50 photons each to have
acceptable counting statistics. The background subtracted
spectrum is very soft and we tried several single-component
spectral models (see Fig.~\ref{fig:spec} and Table
\ref{fit_param}). All models provide a statistically acceptable
fit to the data. However, the power-law index is unplausibly large,
$\Gamma\sim 5.5$, and the Raymond-Smith model converges toward very
low metal abundances (${\rm Z< 10^{-2}\,Z}_\odot$), a pure
bremsstrahlung with a temperature too low for emission from
intracluster gas. If the metal abundance is held fixed at solar, the
fit is not acceptable ($\chi_{red}^2 = 4.7$), ruling out emission
from active stellar coronae.  A fit with the synthetic spectra of
Zampieri et al. (\cite{ztzt95}) for thermal emission from unmagnetized
neutron star hydrogen atmospheres, accreting at low rates, was also attempted.
Using this model, parameterized by the total luminosity in units of the
Eddington luminosity (for a 1.4 $M_\odot$ neutron star),
a distance of $\sim 50$ pc is derived which is likely too
small for the associated column density. The fit with a blackbody is
acceptable and gives ${\rm T}\sim 90$ eV, ${\rm N_H}\sim 4.6\times
10^{20} \ {\rm cm}^{-2}$. These values are in reasonable agreement
with what expected for thermal emission from an isolated neutron
star. Furthermore, the value of the column density derived from the
blackbody fit is the lowest among the fitted models and is well
consistent with the total Galactic absorption in the source direction
($\sim 5\times 10^{20} \ {\rm cm}^{-2}$, Dickey \& Lockman
\cite{dl90}). The 0.1--2.4 keV unabsorbed flux is $8.7\times 10^{-12}
\ {\rm erg \, s^{-1} \, cm^{-2}}$ (Table \ref{fit_param}) and the total
blackbody flux $9.1\times 10^{-12} \ {\rm erg \, s^{-1} \, cm^{-2}}$.

Despite the limited number of counts, a timing analysis has been
attempted.  After correcting the photon arrival times to the solar
system baricenter, no statistically significant periodicity has been
observed with a $3\,\sigma$ upper limit on the pulsed fraction of
$\sim 30\%$ in the 1--100 s period range.
A Kolmogorov-Smirnov test performed on the data set assigns
a probability $<2 \sigma$ to the hypothesis that the source is
variable. The source appears not to be variable also on longer
time-scales ($\sim$ months), as confirmed by the consistency of the
count rate between the ROSAT All Sky Survey and the pointed PSPC
observations.

\begin{figure*}[!htb]
\centerline{\psfig{figure=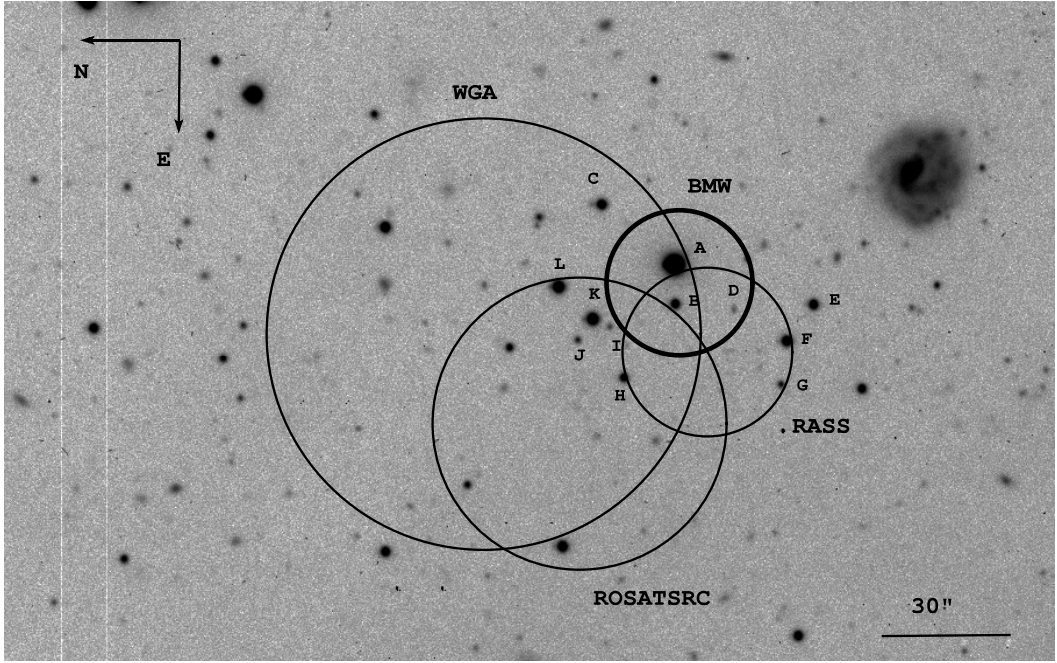,width=14.0cm,angle=0}}
\caption[]{The NTT R field containing RBS 1774. The error circles
refer to the different X-ray positions (see text).
\label{optical}}
\end{figure*}

\subsection{Optical Observations}
\label{opt-obs}

A deep image of the field containing RBS 1774 was taken on 27 May 2001
at the ESO 3.5m New Technology Telescope (NTT) at La Silla. We used
the ESO Multi Mode Instrument (EMMI) with a Tektronix CCD of $2048\times2048$
pixels yielding a field of view of $9.1\arcmin\times 8.6\arcmin$ at a
resolution of 0.27$\arcsec$/pixel. The night was photometric with a
seeing of about 0.8$\arcsec$. Two exposures of 900 s each
were obtained in the red (R) bandpass. The data were reduced using
standard {\sc ESO-MIDAS} procedures for bias subtraction and
flat-field correction. The two images were co-added
and astrometry was performed using the USNO-A2 star catalogue
which provides an absolute positional accuracy of $\sim 0.5\arcsec$
(Monet \cite{mon98}). The central part of the field is shown in
Fig.~\ref{optical}, where the $90\%$ error circles relative to the
different X-ray positions are also shown. As it can be seen, the error
circles partially overlap within the error box provided by
our re-analysis of RASS data. Although this is by no means conclusive,
it lends support to our determination of the source
position. Within our error circle two relatively
bright objects (A and B in Fig.~\ref{optical}) are visible.
The brighter one (A) is the USNO-A2 star
U0900\_19784650, with catalogue magnitudes ${\rm m_R} = 15.5$ and
${\rm m_B} =16.0$. The fainter object is not classified.
The photometric analysis of these two objects
in our NTT image was performed with the Sextractor program (Bertin \& Arnouts
\cite{ber96}). The photometric calibration was obtained from observations of
Landolt standard stars (Landolt \cite{lan92}).
We derived the magnitudes of several stars in the
field around the source position. The two objects within our error
box have ${\rm m_R}=15.40$ and ${\rm m_R}=19.54$, respectively (see Table
\ref{magnitudes}). Another faint object (D) is revealed inside the error
circle at ${\rm m_R}=22.77$.
No other object is visible down to a limiting red magnitude of $\sim 25$,
even if the bright star saturates a sizable fraction of the area.

Optical images of the field in the B, V and R filters were obtained on
29 June 2001 using the 2.5m Nordic Optical Telescope (NOT) at La
Palma. We used the Andalucia Faint Object Spectrograph and Camera
(ALFOSC) with a Loral-Lesser CCD of 2048x2048 pixels yielding a field
of view of 6.5$\arcmin$ at a resolution of 0.188$\arcsec$/pixel. The
night was favorable to perform photometric observations, with a seeing
of 0.8$\arcsec$ (FWHM). Standard data reduction was applied to the
frames using IRAF. The photometric calibration was obtained from
observations of Landolt standard stars (Landolt \cite{lan92}).  The B
and V magnitudes of the field objects in Fig.~\ref{optical} are
given in Table \ref{magnitudes}. The color indices of the two objects
within the BMW error box (A and B) are consistent with those of
intermediate spectral type stars, whereas object C has a marginally
negative V-R and positive B-V. This behavior may be due to an unusual
optical spectrum or to the presence of emission lines in the V band.

\begin{figure}[!htb]
\centerline{\psfig{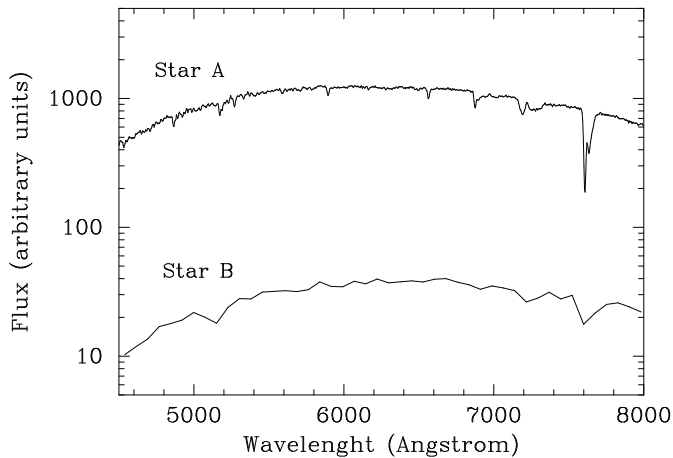}}
\caption[]{The Loiano spectrum of stars A and B. Flux is in arbitrary
units. Strong absorption lines are due to atmospheric absorption.
\label{spectra}}
\end{figure}

Spectroscopic observations of objects A and B were obtained on 24 June
2001 at Loiano Observatory with BFOSC and a seeing of $1.8''$.  We
performed low-resolution (15 \AA, grism \#4) spectroscopy for a total
exposure time of 3000\,s. After applying standard corrections, cosmic
rays were removed and the sky-subtracted stellar spectra were
obtained, corrected for atmospheric extinction. The signal to noise
(S/N) ratio for the dim object was extremely low, providing an
acceptable spectrum only above 4800\AA. No strong emission lines are
observed, testifying that these objects are not Active Galactic Nuclei
(see Fig.~\ref{spectra}). Atmospheric absorption lines (A and B bands)
are clearly visible in the spectra. Star A shows H$\alpha$ and
H$\beta$ absorption lines, pointing to a moderately hot star (class
G). In the case of star B absorption around 5100--5200 \AA \ is seen,
possibly indicating a K main sequence star.  These spectral
classifications are in agreement with color indices derived from the
photometry. The absence of H$\alpha$ emission lines indicates that the
two stars do not have active coronae.

\section{Discussion and Conclusion}
\label{disc}

The soft X-ray source 1RXS J214303.7+065419/RBS 1774 is an optically
unidentified, relatively bright ROSAT source with peculiar
properties. It is interesting to note that Cagnoni et
al. (\cite{cagnoni2001}) have recently performed a systematic search
for bright ROSAT sources with no optical counterpart on the Palomar
Sky Survey from pointed PSPC observations (Blank Field
Sources). However RBS 1774 does not appear in their list because it
does not lie within the inner circle of the PSPC pointing adopted in
their search. This source was probably included by Schwope et
al. (\cite{s99}) among the point-like RBS sources with large ${\rm
f_X/f_V}$ (the small triangle with the lowest hardness ratio in their
Fig.~3).

As discussed in Sect.~\ref{xray-obs}, the X-ray position of the source
reported in the WGA and ROSATSRC catalogues is affected by significant
uncertainties and is better determined from RASS data, that have been
re-analyzed using a wavelet detection algorithm. The two objects
contained within our estimated error box (A and B) have photometric
and spectral properties consistent with those of low-main sequence
stars, with no evidence for coronal activity.
Therefore they are unlikely to be associated with the X-ray
source. A third object (C), which lies just outside the BMW $3\sigma$
error box, has peculiar colors and might not be an ordinary star.
Were it associated with the X-ray source, the inferred X-ray to
optical flux ratio would be $\sim 100$, still consistent with emission
from BL Lac objects, Low Mass X-ray Binaries and Cataclysmic
Variables. However, the X-ray properties of RBS 1774 make these
options unlikely.  In fact, the very soft X-ray spectrum contrasts
with emission from an extragalactic object or an X-ray binary, and the
best-fit blackbody temperature and absence of variability argue
against the identification with a Soft Polar.  Most important,
according to our wavelet detection analysis, the chance that object C
is the counterpart of RBS 1774 is rather slim. No other plausible
optical counterpart appears to be present within the error box, except
possibly for the very faint object D.  At present there are not
sufficient information to establish if this object might actually be
the counterpart of RBS 1774.  Taking as limiting magnitude for the
optical counterpart that of field object D (${\rm m_ R} \sim 22.8$)
sets a lower limit to the X-ray to optical flux ratio of $\sim 1000$.


\begin{table}
\caption{Magnitudes of field objects in the R, V and B bands.
\label{magnitudes}}
\begin{tabular}{cccccc}
        &  R              &         V       &     B    & V--R & B--V \\
\hline
A       & $15.40\pm0.05$  & $15.95\pm 0.05$ & $16.69\pm 0.05$ & 0.6 & 0.7 \\
B       & $19.54\pm0.01$  & $20.0\pm 0.1$   & -- & 0.5 & -- \\
C       & $19.00\pm0.01$  & $18.87\pm 0.05$ & $19.0\pm 0.1$  & -0.1 & 0.1 \\
D       & $22.77\pm0.06$  &       --        & -- & --  & -- \\
E       & $19.38\pm0.01$  & $19.9\pm 0.1$   & -- & 0.5 & -- \\
F       & $19.05\pm0.01$  & $19.4\pm 0.1$   & -- & 0.4 & -- \\
G       & $21.32\pm0.03$  & $21.7\pm 0.1$   & -- & 0.4 & -- \\
H       & $20.10\pm0.01$  & $21.3\pm 0.1$   & -- & 1.2 & -- \\
I       & $22.05\pm0.05$  &       --        & -- & --  & -- \\
J       & $21.56\pm0.03$  & $22.2\pm 0.1$   & -- & 0.6 & -- \\
K       & $18.25\pm0.01$  & $18.96\pm 0.05$ & -- & 0.7 & -- \\
L       & $18.27\pm0.01$  & $19.03\pm 0.05$ & $20.0\pm 0.1$ & 0.8 & 1.0 \\
\end{tabular}

\smallskip
\noindent Star A saturates the NTT R image and we adopt the NOT value.
Estimated errors of 0.05--0.1 mag in the NOT values are due to systematic
errors and uncertainties in the standard calibration.

\end{table}

The very large inferred ${\rm f_X/f_V}$ of RBS 1774 rules out any
known class of X-ray sources but Isolated Neutron Stars. In fact, the
X-ray properties are remarkably similar to those of other INS
candidates detected so far by ROSAT. The soft spectrum (${\rm T}\sim
90$ eV), low column density (${\rm N_H}\sim 5\times 10^{20}$ ${\rm
cm}^{-2}$) and absence of variability in the X-ray data are consistent
with this interpretation. Neglecting intrinsic absorption and
considering a standard value for the average density of the interstellar
medium in the source direction (${\rm n_e} \sim 0.6$ cm$^{-3}$), the
inferred column density allows us to derive a rough estimate for the
source distance of about 300 pc. Given that ${\rm N_H}$ is close to the total
Galactic contribution along the line of sight, this distance should be
considered a lower limit. The source luminosity is ${\rm L} \sim
10^{32} ({\rm d}/300 \, {\rm pc})^2$ erg s$^{-1}$, consistent with
both a young, cooling and a low velocity, accreting neutron star. If
the source is an old, accreting neutron star, the accretion flow may
be in part channelled to the magnetic poles. The absence of modulation
in the X-ray flux at the star spin frequency in the ROSAT data may
simply be due to the low counting statistics and does not rule out
this possibility. In fact, our derived upper limit to the pulsation
amplitude ($\sim 30\%$) is by no means conclusive and a modulation may
well be present with a lower rms amplitude.

The inferred bolometric flux and best fit blackbody temperature make it
possible to estimate the radius of the neutron star as a function of distance. 
Neglecting gravitational redshift, it is: ${\rm R} = 3.2
\, {\rm f}^{-1/2} \gamma^2 \, ({\rm d}/300 \, {\rm pc})$ km. Here ${\rm f}$
is the fraction of the emitting area and $\gamma$ the intrinsic X-ray
spectral hardening factor. The derived upper limit to the pulsation
amplitude implies ${\rm f} \sim 1$. The actual value of $\gamma$
depends on radiative transfer effects across the neutron star
atmosphere and is significantly affected by the chemical composition
and, possibly, the star magnetic field. Given that $\gamma \geq 1$, by
using a realistic value for the neutron star radius it is possible to
derive an upper limit to the source distance of $\sim 900 \, ({\rm
R}/10 \, {\rm km})$ pc. Higher quality X-ray spectra will allow us to
better constraint the spectral distribution and hence the value of
$\gamma$.

In addition to deeper optical observations, an accurate
re-determination of the X-ray position and spectrum, and a systematic
search for periodicities are definitely required to improve our
understanding of RBS 1774. In particular, the unprecedented positional
accuracy and spectral resolution of the instruments on board the
Chandra and XMM satellites provide a unique opportunity to confirm the
identification of RBS 1774 with the seventh known Isolated Neutron
Star.

\begin{acknowledgements}
We thank G.L. Israel for providing us with the software to compute the
limit on the pulsed fraction, E. Bertone for his help in reducing
spectral data, M. Cropper for a helpful discussion on the
properties of Soft Polars and the referee Marten van Kerkwijk for
useful comments. Part of the optical data presented here
have been taken using ALFOSC, which is owned by the Instituto de
Astrofisica de Andalucia (IAA) and operated at the Nordic Optical
Telescope under agreement between IAA and the NBIfAFG of the
Astronomical Observatory of Copenhagen. Work partially supported by
the Italian Ministry for Education, University and Research (MIUR)
under grant COFIN-2000-MM02C71842.
\end{acknowledgements}


\begin{thebibliography}{}

\bibitem[1996]{ber96}
Bertin, E., \& Arnouts, S. 1996, A\&AS, 117, 393
\bibitem[1987]{bre87}
Bregoli, G., Federici, L., Merighi, R., et al. 1987, in
ESO--OHP  Workshop on the Optimization of the Use of CCD Detectors in
Astronomy Proc. A88--13301 03--89 (Garching: ESO), 177
\bibitem[2001]{cagnoni2001}
Cagnoni, I., Celotti, A., Elvis, M., Kim, D.-W., \& Nicastro, F.
2001, in Proc.of the Fourth Italian Conference on AGNs (Mem. SAIt),
in press (astro-ph/0006257)
\bibitem[1999]{cam99}
Campana, S., Lazzati, D., Panzera, M.R., \& Tagliaferri, G.
1999, ApJ, 524, 423
\bibitem[1990]{dl90}
Dickey, J., \& Lockman, F. 1990, ARA\&A, 28, 215
\bibitem[1997]{hetal97}
Haberl, F., Motch, C., Buckley, D.A.H., Zickgraf, F.-J., \& Pietsch, W.
1997, A\&A, 326, 662
\bibitem[1999]{hetal99}
Haberl, F., Pietsch, W., \& Motch, C. 1999, A\&A, 351, L53
\bibitem[1998]{hh98}
Heyl, J.S., \& Hernquist, L. 1998, MNRAS, 297, L69
\bibitem[1998]{hk98}
Heyl, J.S., \& Kulkarni, S.R. 1998, ApJ, 506, L61
\bibitem[1998]{kk98}
Kulkarni, S.R., \& van Kerkwijk, M.H. 1998, ApJ, 507, L49
\bibitem[1992]{lan92}
Landolt, A.U. 1992, AJ, 104, 340
\bibitem[1999]{laz99}
Lazzati, D., Campana, S., Rosati, P., Panzera, M.R., \& Tagliaferri, G.
1999, ApJ, 524, 414
\bibitem[1998]{mon98}
Monet, D.G. 1998, AAS Meeting \#193, \#120.03
\bibitem[1999]{mo99}
Motch, C., Haberl, F., Zickgraf, F.-J., Hasinger, G., \& Schwope, A.D.
1999, A\&A, 351, 177
\bibitem[2000]{m2000}
Motch, C. 2000, in Bologna X-ray Astronomy 1999, in press
(astro-ph/0008485)
\bibitem[1998]{mh98}
Motch, C., \& Haberl, F. 1998, A\&A, 333, L59
\bibitem[1999]{nt99}
Ne\"uhauser, R., \& Tr\"umper, J.E. 1999, A\&A, 343, 151
\bibitem[2000]{p2000}
Popov, S.B., Colpi, M., Prokhorov, M.E., Treves, A., \& Turolla, R.
2000, ApJ, 544, L53
\bibitem[1999]{s99}
Schwope, A.D., Hasinger, G., Schwarz, R., Haberl, F., \& Schmidt, M.
1999, A\&A, 341, L51
\bibitem[2000]{s00}
Schwope, A.D., Hasinger, G., Lehmann, I., et al. 2000, AN, 321, 1
\bibitem[1987]{ste87}
Stetson, P.B. 1987, PASP, 99, 191
\bibitem[1991]{sto91}
Stocke, J.T., Morris, S.L., Gioia, I.M., et al. 1991, ApJS, 76, 813
\bibitem[2000]{t2000}
Treves, A., Turolla, R., Zane, S., \& Colpi, M. 2000, PASP, 112, 297
\bibitem[2001]{kvk2001}
van Kerkwijk, M.H., \& Kulkarni, S.R., 2001 A\&A, in press (astro-ph/0106265)
\bibitem[1996]{vetal96}
Voges, W., Aschenbach, B., Boller, T., et al. 1996, IAU Circ. 6420
\bibitem[2001]{w2001}
Walter, F.M. 2001, ApJ, 549, 433
\bibitem[1997]{wm97}
Walter, F.M., \& Matthews, L.D. 1997, Nature, 389, 358
\bibitem[1994]{wga94}
White, N.E., Giommi, P., \& Angelini, L. 1994, IAU Circ. 6100
(http://wgacat.gsfc.nasa.gov)
\bibitem[1995]{ztzt95}
Zampieri, L., Turolla, R., Zane, S., \& Treves, A. 1995, ApJ, 439, 849
\bibitem[1994]{zim94}
Zimmermann, H.-U. 1994, IAU Circ. 6102

\end{thebibliography}
\end{document}